# Cornell Caltech Atacama Telescope (CCAT): a 25 m aperture telescope above 5000 m altitude

Thomas A. Sebring[1], Riccardo Giovanelli[1], Simon Radford[2], Jonas Zmuidzinas[2]
[1]Center for Radiophysics and Space Research, Cornell University, Ithaca, NY 14853
[2]California Institute of Technology, Pasadena, CA 91125

## ABSTRACT

Cornell, California Institute of Technology (Caltech), and Jet Propulsion Lab (JPL) have joined together to study development of a 25 meter sub-millimeter telescope (CCAT) on a high peak in the Atacama region of northern Chile, where the atmosphere is so dry as to permit observation at wavelengths as short as 200 micron. The telescope is designed to deliver high efficiency images at that wavelength with a total ½ wavefront error of about 10 microns.  With a 20 arc min field of view, CCAT will be able to accommodate large format bolometer arrays and will excel at carrying out surveys as well as resolving structures to the 2 arc sec. resolution level.  The telescope will be an ideal complement to ALMA.  Initial instrumentation will include both a wide field bolometer camera and a medium resolution spectrograph.  Studies of the major telescope subsystems have been performed as part of an initial Feasibility Concept Study.  Novel aspects of the telescope design include kinematic mounting and active positioning of primary mirror segments, high bandwidth secondary mirror segment motion control for chopping, a Calotte style dome of 50 meter diameter, a mount capable of efficient scanning modes of operation, and some new approaches to panel manufacture.  Analysis of telescope performance and of key subsystems will be presented to illustrate the technical feasibility and pragmatic cost of CCAT.  Project plans include an Engineering Concept Design phase followed by detailed design and development.  First Light is planned for early 2012.

## 1. INTRODUCTION

Cornell University, the California Institute for Technology, and the Jet Propulsion Laboratory are jointly studying the construction of a 25 m telescope for submillimeter astronomy on a high mountain in northern Chile. Observations with this Cornell Caltech Atacama Telescope (CCAT) will address fundamental themes in contemporary astronomy, notably the formation and evolution of galaxies, the nature of the dark matter and dark energy that comprise most of the content of the universe, the formation of stars and planets, the conditions in circumstellar disks, and the conditions during the early history of the Solar system. With a 20 arc min field of view, the CCAT will emphasize wide field mapping, complementing the narrow field, high resolution capabilities of ALMA. The CCAT conceptual design and feasibility study was completed at the end of 2005 and the project received a strong endorsement from an independent review committee in January, 2006. The CCAT partnership is now expanding and the project plan aims for first light in early 2012.

## 2. BACKGROUND

Over the past decade, Cornell, Caltech, and other groups have evaluated conditions for submillimeter astronomy at sites at and above 5000 m in the Atacama region of Chile. The measurements demonstrate these sites enjoy excellent observing conditions with extremely low water vapor content. Observing conditions are considerably better than Mauna Kea and are comparable to the South Pole. As a result several telescopes have already been established in the area, notably the ALMA now under construction on a 5000 m plateau near the village of San Pedro de Atacama. Furthermore, conditions on the peaks surrounding the ALMA site are even better, particularly when thermal inversions trap much of the water vapor below the mountain summits (Figure 1).

This presents the opportunity for development of a large telescope for submillimeter astronomy, the CCAT. The science objectives emphasize wide field observations with large format bolometer cameras. CCAT has been seen as complementary to ALMA. Although ALMA's long baselines will provide unequalled resolution, its small instantaneous field of view severely limits its wide field survey capability. CCAT, on the other hand, will provide wide field images with an abundance of sources for later follow up by ALMA. Over the past two years, Cornell, Caltech, and JPL have undertaken a concept design feasibility study to evaluate the telescope design performance, feasibility, and cost. The science objectives determined the configuration and performance requirements. The study was funded by the three institutions and was performed by members of the institutions and contractors under the leadership of a Project Manager and Deputy. This paper summarizes the study results and project status.

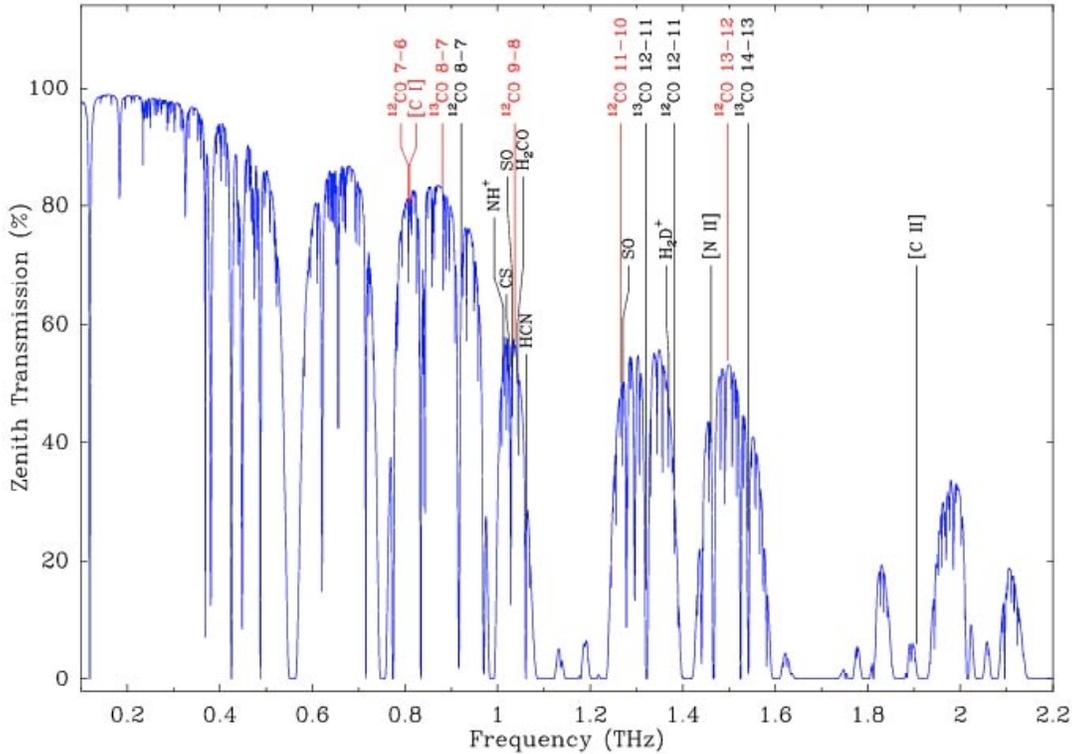

**Figure 1.** Atmospheric transmission measured on 2005 January 24 with an FTS at the 5500 m site of the Smithsonian RLT on Cerro Sairecabur, about 40 km north of the 5000 m ALMA site. The inferred water vapor column depth is only 93 μm (Marrone et al. 2005 astro-ph/0505273).

## 3. SCIENCE

As well as being substantially larger and more sensitive than existing submillimeter telescopes, CCAT will be the first large submillimeter telescope designed specifically for wide field imaging. Hence it will provide an unparalleled ability to address key astronomical questions by mapping large areas of the sky. The scientific priorities include:
- How did the first galaxies form?

CCAT will detect hundreds of thousands of primeval galaxies from the era of galaxy formation and assembly, $z =$ 2–4 or about 10–12 billion years ago, providing for the first time a complete picture of this process. In addition, CCAT will probe the earliest bursts of dusty star formation as far back as $z \sim 10$, less than 500 million years after the Big Bang when the Universe was ~ 4% of its current age.
- What is the nature of the dark matter and dark energy that fill the Universe?

CCAT will image the Sunyaev-Zel'dovich effect in hundreds of clusters of galaxies selected from current and planned southern-hemisphere cluster searches. These images will be important for understanding how clusters form and evolve and for the interpretation of the survey data to constrain crucial cosmological parameters, i. e., $\Omega M$, $\Omega \Lambda$, and the dark energy equation of state, independently of other techniques such as Type Ia supernova and direct CMB measurements.
- How do stars form?

By surveying molecular clouds in our Galaxy, CCAT will detect cold dense cores that collapse to form stars, providing for the first time a complete census of star formation down to very low masses. In nearby molecular clouds, CCAT will be able to detect cold cores much smaller than the lowest mass stars (0.08 $M_\odot$).

• How do conditions in circumstellar disks determine the nature of planetary systems and the possibilities for life?
CCAT will image dust produced by collisional grinding of planetesimals around other stars to allow determination of the dynamical effects of planets on the dust distribution and, hence, the properties of planetary orbits. In concert with ALMA, CCAT will study disk evolution from early, Class I, to late, debris disk, stages.

• How did the Solar System form?
Beyond Neptune, the Kuiper belt is a relic containing a record of the processes that operated in the early solar system, i.e., the accretion, migration, and clearing phases. CCAT will determine sizes and albedos for hundreds of Kuiper belt objects, thereby providing information to anchor models of planet formation in the early solar system.

Although hydrogen and helium make up over 98% of the baryonic matter in the Universe, in many cases the heavier elements, notably carbon, oxygen, silicon, and iron, allow us to discover and study distant objects. These elements form complex molecules and small dust particles that in many astrophysical environments obscure optical and ultraviolet photons and radiate predominantly at submillimeter wavelengths. Many of the most powerful and interesting phenomena in the universe, from star forming regions in our own galaxy to entire galaxies, are shrouded by dust and are completely inaccessible with optical observations. This makes the submillimeter a particularly valuable probe of many astrophysical sources.

## 4. TELESCOPE REQUIREMENTS

Table 1 illustrates the top-level requirements for CCAT determined from the science objectives.

**Table 1.** Summary of CCAT performance requirements and goals.

|  | Requirement | Goal | Remarks |
|---|---|---|---|
| Wavelength range | 350 – 1400 µm | 200 – 3500 µm | |
| Aperture | 25 m | | |
| Field of view | 10' | 20' | |
| Half Wavefront Error | < 12.5 µm rms | < 9.5 µm rms | |
| Site conditions | < 1 mm | < 0.7 mm | Median pwv |
| Polarization | 0.2% | 0.05% | After calibration |
| Emissivity | < 10% @ > 300 µm | < 5% @ > 800 µm | |
| | < 20% @  200 µm | | |
| Pointing, blind | 2" | 0.5" | rms |
| offset | 0.3" | 0.2" | within 1° |
| Repeatability, one hr. | 0.3" | 0.2" | rms |
| Elevation range | 15–90° | | from horizon |
| Azimuth range | ± 270° | | from north |
| Scanning rate | 0.2° s$^{-1}$ | 1° s$^{-1}$ | slow and fast modes |
| acceleration | 0.4° s$^{-2}$ | 2° s$^{-2}$ | rms |
| point. know. | 0.2" | 0.1" | |
| Secondary nutation | ±2.5' @ 1 Hz | | azimuth only |

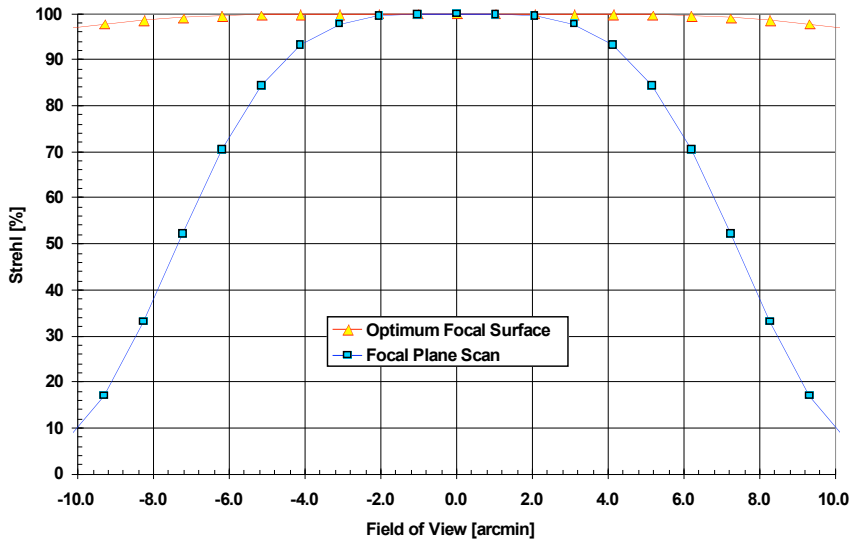

**Figure 2.** Strehl ratio variation across the field of view both on the optimal (curved) focal surface and on a flat focal plane for observations at 200 μm (G. Cortés-Medellin).

## 5. OPTICAL DESIGN

A variety of optical designs and configurations was examined. Ultimately a Ritchey Chrétien design with Nasmyth foci was chosen. This was based on the overall performance over the full desired field of view and the need for rapid changes between multiple instruments. Arrangements with the tertiary mirror both in front of and behind the primary were investigated. Considerations such as secondary mirror size, available space envelope for instruments, and balance of the elevation stage led to the current configuration. Although the study design has an $f/0.6$ primary mirror, one conclusion of the study review is to make the telescope faster and an $f/0.4$ primary is now under consideration. This will reduce the sizes of the dome, the secondary support structure, and the secondary mirror significantly. The optical performance (Figure 2) demonstrates excellent image quality across the field of view. Short wavelength, wide field instruments can accommodate the focal plane curvature either with corrective optics or with a segmented detector array.

## 6. TELESCOPE SITE

Of the several peaks in reasonably close proximity to ALMA, Cerro Chajnantor (5600 m) has been selected as the baseline CCAT site. Other projects are also interested in this mountain, which lies within the recently expanded CONICYT science preserve, and the University of Tokyo has constructed a road to the summit area. For CCAT, the preferred position is a small plateau about 150 m northeast and 50 m below the summit ridge (Figure 3). This location is shielded from the prevailing westerly winds and is out of view of the village of San Pedro de Atacama. A preliminary geotechnical assessment suggests there is a bearing stratum of competent rock, which would allow fairly efficient foundations for both the facilities and the telescope mount. Additional measurements of meteorological and observing conditions and geotechnical characteristics will be performed during the coming year.

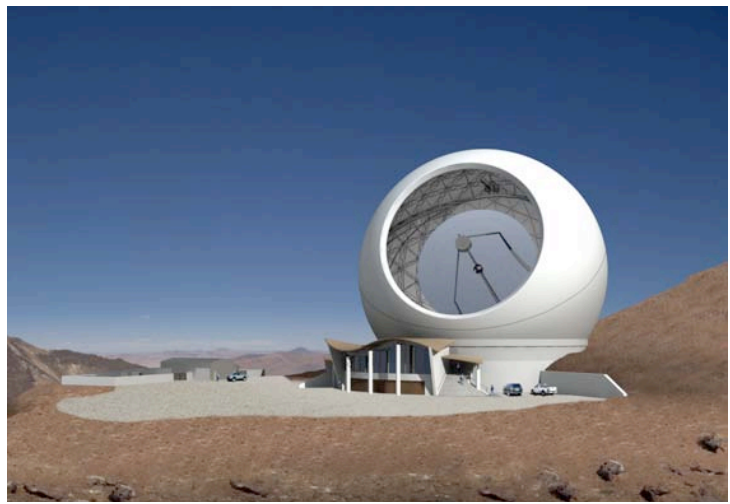

**Figure 3.** Concept view of CCAT at 5600 m on Cerro Chajnantor, Chile (M3 Engineering).

## 7. FACILITY DESIGN

Under contract, M3 Engineering (Tucson, AZ) developed a concept design of the telescope facility (Figure 4) including a control room, an office area, a computer and electronics room, and an instrument preparation lab. The concept provides an early indication of the approach to foundations. Particular attention was paid to construction approaches that minimize on-site work, for example, the use of pre-cast concrete panels as building elements. Much of the building interior is oxygen enriched to enhance the safety, comfort and productivity of observers and personnel engaged in work on instruments. Portable oxygen will be used for exterior work during operations and during all phases of construction. Separate buildings house equipment such as chillers, compressors, etc., so vibration and heat loading on the telescope are minimized. M3 also developed a design for a lower altitude support facility near San Pedro de Atacama. This base facility provides housing for operations personnel as well as a remote control room, offices, and instrument workshops.

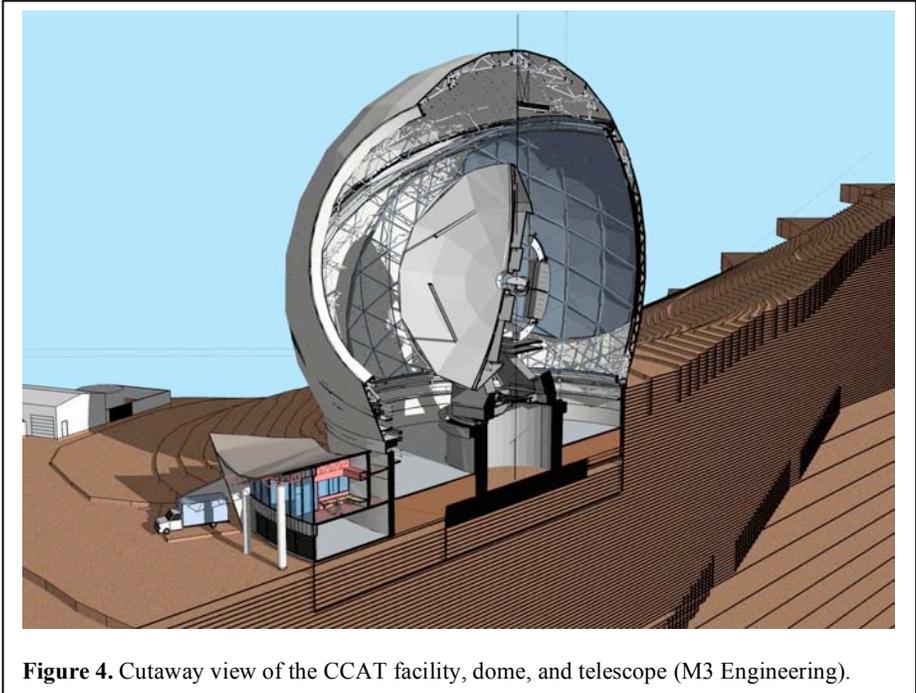

**Figure 4.** Cutaway view of the CCAT facility, dome, and telescope (M3 Engineering).

## 8. TELESCOPE DOME

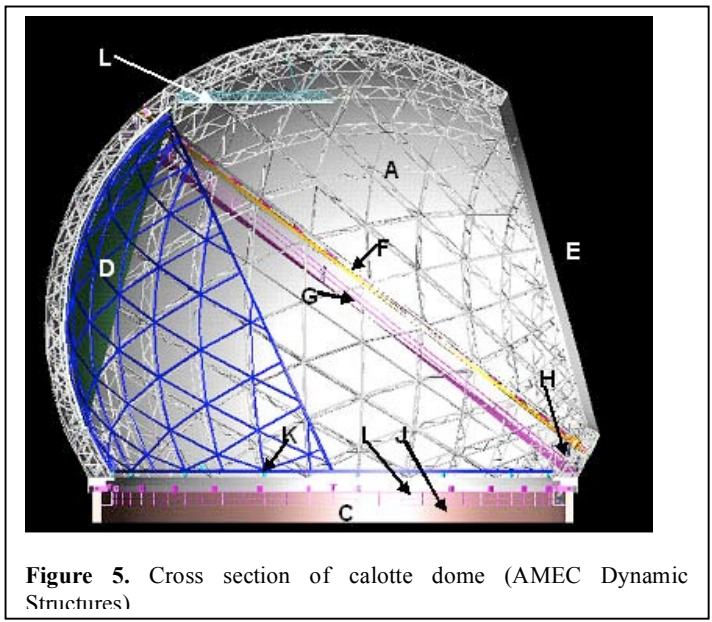

**Figure 5.** Cross section of calotte dome (AMEC Dynamic Structures)

AMEC Dynamic Structures (Vancouver, BC) were hired to develop concept designs for the dome. It was concluded a dome would be required based on measured wind speeds and analysis of effects of wind loading on panel alignment and pointing. The study of domes began with parametric assessment of several dome types. FEM analysis indicated the Calotte type (Figure 5) chosen for concept development weighed significantly less than other types, exhibited less and more uniformly distributed stress throughout the structure, and could be completely balanced in operation. The results indicated that this approach would cost less to build as well as to operate. The structure of the dome in this concept is a regular rib and tie structure of 1 m deep welded trusses. Either this approach or a geodesic concept will be further developed as the Project progresses. The advantage of the regularized structure is lower manufacturing cost, smaller shipping packages, and simple bolted construction on site with minimum alignment or field welding required. The dome has a diameter of 50 m at the equator, but this may be reduced to ~40 m as a consequence of adopting a faster primary focal ratio. Design of the azimuth rotation stage is fairly straightforward and follows successful practice of many large astronomical telescope domes. The tilted rotation stage (item F in Figure 5) is more challenging and is probably the crux engineering issue in development of the dome. A third rotating stage (item D in Figure 5) is the shutter assembly, which can be rotated independently of the azimuth to bring the paneled shutter into alignment with the aperture. The dome and telescope can rotate independently allowing full work on the telescope within the enclosure. Consideration of aluminum versus steel construction and other design issues will be addressed in the next phase of the Project.

## 9. TELESCOPE MOUNT

The telescope mount layout was developed in parallel with the optical design. Variations in elevation axis position and instrument locations were traded against secondary mirror size and overall optical design. A fundamental requirement was to accommodate the full 20 arc minute field of view by allowing a 1 m diameter beam to pass through the elevation bearings. The instruments were placed inside the primary mirror outer diameter to reduce the the secondary mirror size, the back focal distance, and the dome size. The concept mount design (Figure 6), developed by VertexRSI (Richardson, TX), employs hydrostatic bearings for azimuth, where the loads are high. The journals mount to the cylindrical pier and radial runout is controlled by use of a central pintle bearing. Two diametrically opposed pairs of motors and gearboxes drive against a large helical gear mounted to the inside diameter of the azimuth journal. The much lighter elevation stage uses commercially available rolling element bearings and a pair of geared motors driving a helical gear sector following common radiotelescope practice. Encoders for both axes are Heidenhain tape encoders with multiple read heads. Preliminary servo control models were built and exercised, indicating that the telescope can meet stringent requirements for pointing precision and scanning motions. Future design work on the mount will include optimizing the M2 support structure for minimum obscuration, structural optimization, weight reduction, and modularization.

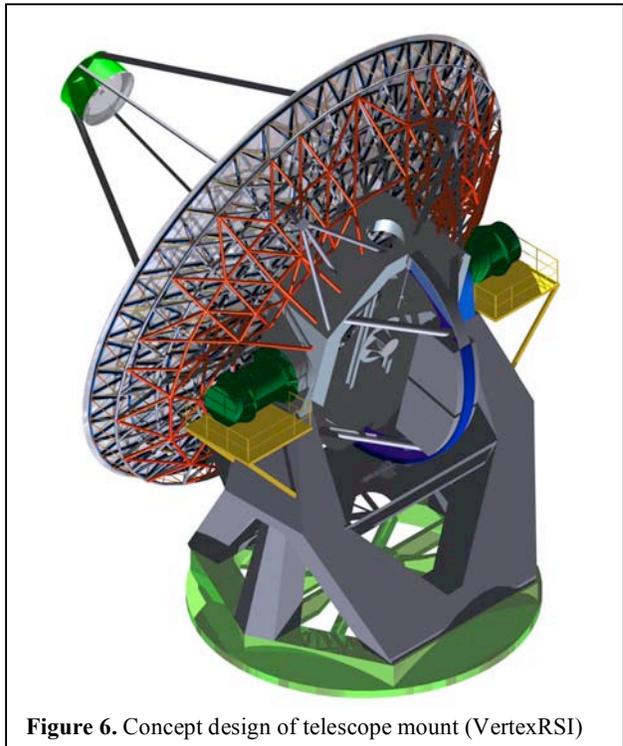

**Figure 6.** Concept design of telescope mount (VertexRSI)

Manufacture of the mount to best accommodate trial assembly, packaging and shipping, and on-site integration will be an integral part of ongoing design efforts.

## 10. PRIMARY MIRROR

Following investigation of the successful approaches used on both optical and radio wavelength segmented primary mirrors, it was determined that the most fruitful approach for investigation was rings of replicated panels. This approach has the benefit of having only 6 or 7 types of panels with maximum edge dimensions of from 2 to 1.8 m, respectively. The baseline segmentation pattern has 187 keystone panels in 7 rings with characteristic sizes of 1.84 m (Figure 7). Contractors investigated three different panel technologies: CFRP/Al honeycomb laminates (CMA, Tucson, AZ), precision molded lightweight borosilicate (ITT, Rochester, NY), and lightweight cored SiC (Xinetics, Devens, MA). The panel manufacturers were asked to design panels, analyze structural performance, and postulate a manufacturing process. Both CFRP/Al and precision molded glass appear feasible and additional development will be undertaken. The overall surface error tolerance for the panels, including thermal and gravitational deformation, is 5 μm rms (Figure 8). The panels meet this requirement when kinematically mounted on just three points, eliminating the need for complex multi-point whiffle tree type supports. Analysis verified that either panel technology can meet this requirement.

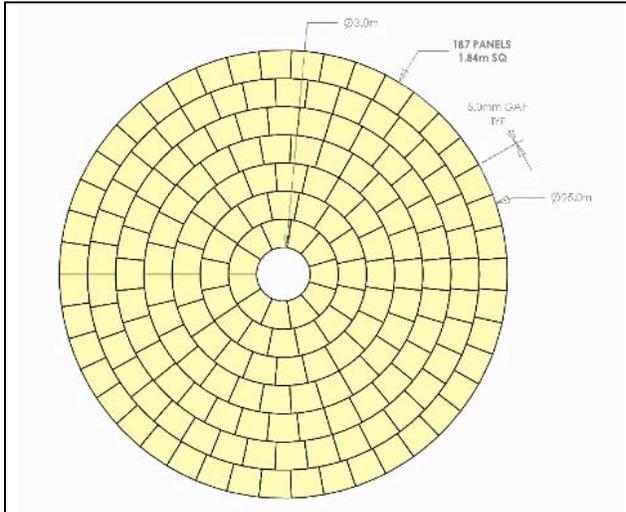

**Figure 7.** Primary mirror segmentation pattern.

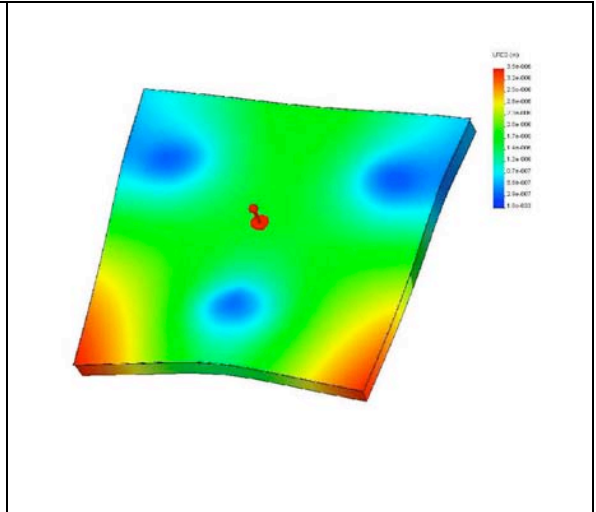

**Figure 8.** Gravitational deformations of a CFRP/Al mirror panel (CMA). Scale is 0 to 3.5 μm.

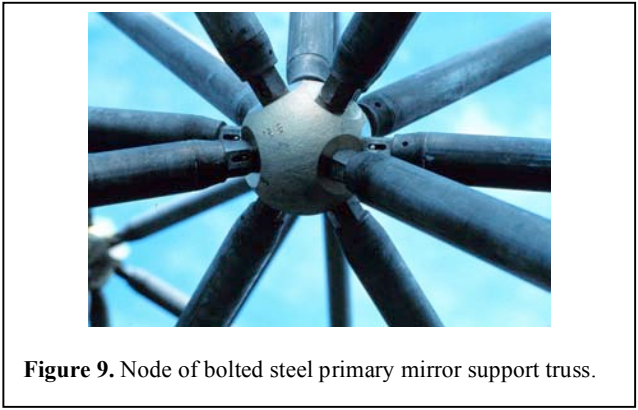

**Figure 9.** Node of bolted steel primary mirror support truss.

A bolted steel truss will support the primary mirror. This approach provides adequate structural characteristics in an inexpensive truss that is easily shipped and assembled on site. Although CFRP is more dimensionally stable than steel, a 25 m diameter CFRP truss would be prohibitively expensive, about seven times more than steel. With a steel truss, active panel positioning is required so the mirror segments are mounted on actuators. Closed loop active surface control has been successfully achieved in current optical segmented telescopes and an open loop active surface has been implemented on the Caltech Submillimeter Observatory (CSO).

## 11. ACTIVE PRIMARY SENSING AND CONTROL

The overall primary mirror surface and the segment positions will be calibrated with shearing interferometry observations of planets. This technique has been successfully used at the CSO. Maintenance of the segment alignment will largely rely on edge sensors and commercial suppliers have proposed systems that are reasonable extrapolations of the sensors used on the Hobby Eberly Telescope and the Southern African Large Telescope. Edge sensors alone are insufficient, however, because they do not sense well such low order deformation modes as the global radius of curvature. Supplemental metrology to resolve these ambiguities may include a laser system under development at JPL for segmented optics space missions. Figure 10 illustrates a notional sparse distribution of these laser sensors to supplement conventional edge sensors. Alternative supplemental sensors have also been examined, including a Shack Hartmann approach using a point source and small return mirrors and an IR wavefront sensing guider using stellar sources. This last approach requires optical surfaces specular at the sensing wavelength, but this appears achievable via the panel replication techniques studied at little additional cost.

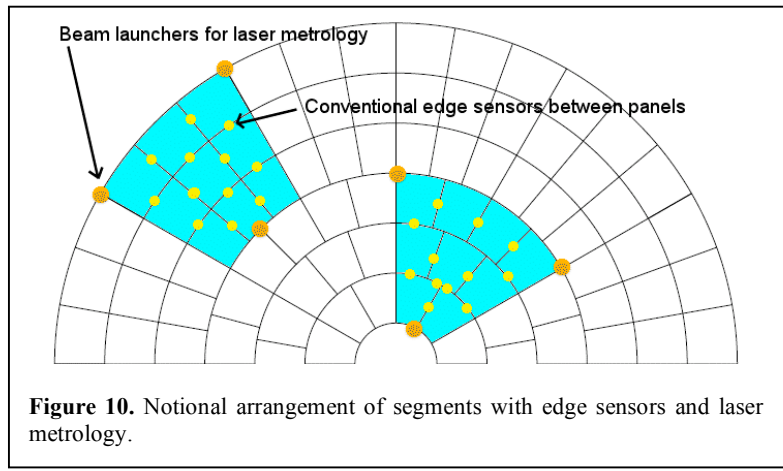

**Figure 10.** Notional arrangement of segments with edge sensors and laser metrology.

## 12. SECONDARY AND TERTIARY MIRRORS

With an $f$/0.6 primary, the secondary mirror is 3.3 m in diameter. This size, combined with the requirement of extremely light weight for nutation and telescope scanning modes, inspired a novel concept for a segmented and hi-bandwidth active M2 system (Figure 11). In this approach four identical mirror segments are made using the same replication process selected for primary mirror segments. Each lightweight segment is mounted with a high-bandwidth hexapod to a CFRP structure using a reactionless design. The panels can then be positioned independently to form the parent optic, but also moved in concert at bandwidth and with sufficient precision to yield nutation of the integrated surface while maintaining alignment. This approach appears feasible based on prior successful high-bandwidth hexapod systems built by the study contractor, CSA Engineering (Mountain View, CA). In the next phase of CCAT development we will consider a change to make the telescope $f$/0.4. In addition to other changes, this reduces M2 to about 2.6 m diameter, a more reasonable size to manufacture a monolithic lightweight M2 and hence employ a more conventional nutator. The tertiary mirror rotates to direct the image to either Nasmyth focus or to the bent Cassegrain foci, which can support smaller instruments, such as calibration instrumentation.

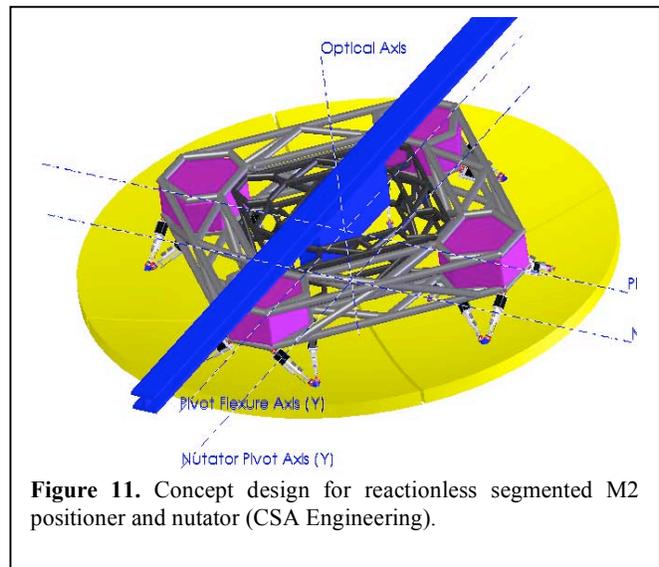

**Figure 11.** Concept design for reactionless segmented M2 positioner and nutator (CSA Engineering).

## 13. SCIENCE INSTRUMENTATION

Because the primary science objectives of the CCAT emphasize wide field imaging and surveys, the first light instruments will be two cameras: a short wavelength camera (SWCam) for submillimeter wavelengths (200–620 µm) and long wavelength camera (LWCam) for near mm wavelengths (750–2000 µm). Both technology and performance constraints make it more practical to build two cameras rather than one combination instrument. To reduce the operations support demands, both cameras will use closed cycle cryogenic systems consisting of pulse tube coolers followed by $^4$He and $^3$He or ADR stages. The SWCam has 32,000 directly illuminated TES silicon bolometers spaced to Nyquist sample a 5′ × 5′ field of view (FoV) at 350 µm. Mesh filters, well matched to the atmospheric windows and mounted in a wheel immediately behind the Lyot stop, select the observing wavelength.

The LWCam uses an array of slot dipole antenna coupled bolometers. Microstrip bandpass filters separate the frequency bands, providing simultaneous multicolor observations. Both TES and MKID bolometers are under consideration. Plate scales and pixel counts are wavelength dependent, covering the entire 20′ × 20′ FoV at the longest wavelengths. Reflective coupling optics are employed to insure minimum optical loading. These cameras are challenging, but achievable through modest advances in the current array technologies. Existing, previous generation instruments, including direct detection and heterodyne spectrometers developed for other facilities, will be brought to CCAT to enhance the scientific yield of the two cameras. Although these existing instruments cannot accomplish all the primary CCAT science objectives, they will provide important supplementary capabilities, especially in the early years of operation. Looking to the future, foreseeable instrument developments will extend the CCAT science return for many years.

## 14. SCHEDULE AND BUDGET

To achieve maximum synergy with ALMA, the CCAT development schedule (Table 2) aims to bring the telescope into operation in 2014, near the start of full ALMA operations. The estimated cost of CCAT is $100 M including $20 M for the two first light instruments.

**Table 2.** CCAT Project Schedule

| | |
|---|---|
| Concept/Feasibility Study | 2005 |
| Study Review | 2006 January |
| Engineering Concept Design | mid 2006 – mid 2007 |
| Concept Design Review | 2007 August |
| Development | mid 2007 – 2010 Q4 |
| General Construction | 2008 – 2010 |
| Integration | 2010 – 2012 |
| First Light | 2012 Q1 |
| Commissioning | 2012 – 2013 |
| Completion | 2014 Q1 |

## 15. SUMMARY

CCAT is a significant project for addressing key astronomical questions by mapping large areas of the sky at submillimeter wavelengths. A high, dry site affords excellent observing conditions. The combination of large aperture and wide field will be unequalled. The CCAT provides a platform for instruments that will take maximum advantage of rapidly developing detector technology. CCAT's proposed technologies are based on successful prior telescopes and are largely within the state of the art. Several other institutions have expressed significant interest in joining the CCAT Project. We anticipate development will continue at a pace consistent with the proposed schedule.